\documentclass{iaus}
\usepackage{graphicx}

\title[Bars driven by the Cosmology] 
{Bars driven by the Cosmology in stellar-gaseous disks}

\author[Anna Curir, Paola Mazzei, Giuseppe Murante]   
{Anna Curir$^1$
 Paola Mazzei $^2$  and Giuseppe Murante$^1$}

\affiliation{$^1$Astronomical Observatory of Torino \\ Strada Osservatorio 20,
10025, Pino Torinese (Torino), Italy \\ email: {\tt curir@oato.inaf.it} \\[\affilskip]
$^2$Astronomical Observatory of Padova, \\ vicolo Osservatorio 5,
35122, Padova, Italy \\email: {\tt paola.mazzei@oapd.inaf.it}}

\pubyear{2008}
\volume{254}  
\pagerange{1-6}
\jname{The Galaxy Disk in Cosmological Context}
\editors{J. Andersen, J. Bland-Hawthorn \& B. Nordström, eds.}
\begin{document}

\maketitle

\begin{abstract}

We  present  the first attempt to
analyse the growth of the bar instability in  stellar-gaseous disks 
evolving in a  fully consistent cosmological scenario. 
We explored the role of the cosmology on pure stellar disks with different mass embedded 
in a cosmological dark matter halo.
We deepened such a study by analysing the
impact of different gas fractions  and of the star formation onset.

We found that in all these cases, the stellar bar arising inside the less massive 
disks,  i.e., dark matter (DM)-dominated disks, is still
living at redshift zero even if the gas fraction 
exceeds half of the disk mass. Such a bar is a genuine product of the cosmology.
However,  in the most massive disks there is  a  threshold value for their gas
percentage and 
lower limit for the central gas concentration able to destroy the bar when the
star formation rate is switched off.

On the other hand in the simulations with star formation
the central mass concentration of gas and of the new stars has a mild action
on the ellipticity of the bar but is not able to destroy it; at z=0 the
stellar bar strength is enhanced by the star formation.
Even if our results  qualitatively agree with the classical ones, i.e. with  criteria concerning bar instability derived 
outside the cosmological framework,
the same  criteria cannot be validated for the DM-dominated disks.

\end{abstract}

\firstsection 
\section{Introduction}

In a pioneering paper \cite[Curir \& Mazzei (1999)]{Cu99} put the first question mark on the effect of the
halo triaxiality on the bar growth and evolution, by carrying on  SPH simulations of a
disk inside a non spherical halo, in an
isolated context.
They embebbed a stellar-gaseous disk in dark matter (DM)
halos of different masses, shapes and dynamical states. These models allowed
to follow for the first time  the effect of a live
triaxial  halo on the disk evolution. Their first result is that a massive halo not yet 
relaxed has a major role in triggering bar instability.

The gas behaviour in disk galaxies and its connections with the bar feature
has been
studied in several papers  
(\cite[Friedli \& Benz (1993)]{friebe93}, \cite[Berentzen at a. (2001)]{ber01}, \cite[Bourneaud et al. 2005]{Bou05}, \cite[Michel-Dansac \& Wozniak, 2004]{MiWo}).
In all these works 
the evolution of the disk or of the disk-halo simulated system arises in an isolated framework,
outside the cosmological scenario.

The connection between the bar feature and the star formation  process has
already been pointed out in the past. \cite[Martinet \& Friedli
(1977)]{mart77} in particular  observed that
non interacting galaxies  displaying the highest star forming activity have
strong and long bars.
\cite[Mazzei \& Curir (2001)]{Ma01} 
investigated  further the role of a live
non axisymmetric DM halo, with different geometry and dynamical state, on the
star formation and on its role in  the bar triggering.
They showed that the star formation  enhances the bar instability and lengthens the life time of the bar.
However in these works the evolution of the disk+halo system arises in a isolated framework.

In a series of papers  (\cite[Curir et al.  (2006)]{Cu06}, \cite[Curir et
al. (2007)]{Cu07}, \cite[Curir et al. (2008)]{Cu08}) we investigated for the first time the
 growth of the
bar instability in a cosmological context.  
For this purpose we followed the behaviour  of  an exponential baryonic  disk
inserted  in a  cosmological DM halo evolving in a fully cosmological framework.
Our model cannot be viewed as a general galaxy evolution model, since the
gradual formation and growth of the stellar disk has not been taken into
account. 
 Thus our work cannot be compared with the more recent papers by \cite[Governato et al. (2004)]{Gov04}, \cite[Robertson et al. (2004)]{Spri04}, \cite[Sommer-Larsen (2006)]{SLres}, where
the formation of a spiral galaxy within the hierarchical scenario of
structure formation in {$\Lambda$ }-dominated cosmologies has been followed
 self-consistently. 
However, our approach has the advantage of allowing to vary parameters like the
disk-to-halo mass ratio, the gas fraction inside the disk, to switch off and
on the star formation with the aim  to 
analyse the growth of the bar instability and 
its dependence on these parameters  in a self-consistent cosmological framework. 
We embedded a purely stellar  and a  stellar-gaseous disk inside a  cosmological halo selected in a
suitable slice of the Universe (see Fig. 1) and follow its  evolution inside a cosmological
framework: a $\Lambda$CDM  model with
$\Omega_{m}=$0.3, $\Omega_{\Lambda}=$0.7, $\sigma_8=$0.9 and 
$h=$0.7, where $\Omega_{m}$ is
the total matter of the Universe, $\Omega_{\Lambda}$ the
cosmological constant, $\sigma_8$  the normalisation of the power spectrum,  
and $h$ the value of the Hubble constant in units of 100$h^{-1}$
km\,s$^{-1}$Mpc$^{-1}$. 
\begin{figure}
\begin{center}
\includegraphics[width=3.4in]{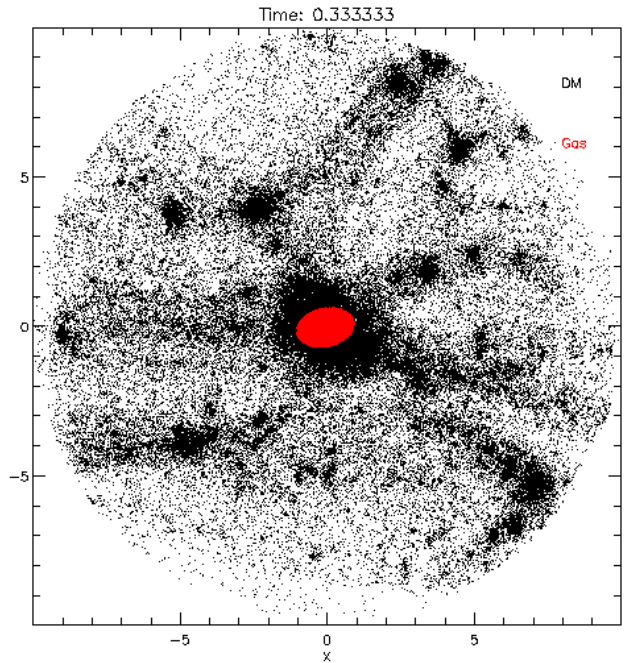}
\end{center}
\caption{The disk (red points) embedded in the DM Cosmology
}
\label{disk}
\end{figure}
\section{Overview}

\subsection{The DM halo}

To select the DM halo, we perform a low-resolution simulation of a {\it
concordance} $\Lambda$ CDM cosmological model, starting from redshift 20.\\
With a standard 'friends of friends algorithm' we select one suitable DM
halo with a mass M\( \sim \)10\( ^{11} \)\( h^{-1} \) M\( _{\odot } \) (at
z=0). We resample it with a multi-mass technique. The particles of the DM halo, and those belonging to a sphere
with a radius $4 h^{-1}$ Mpc, are followed to their Lagrangian position and
re-sampled to an equivalent resolution of 1024 \(^{3} \) particles.  The total
number of DM particles in the high resolution corresponds to a DM mass
resolution of $ 1.21\, 10^{6} h^{-1}\,M_{\odot }$.  The high-resolution DM
halo is followed to the redshift z=0.  We run the DM simulation, to extract
the halo properties in absence of any embedded disk.  The mass of our
halo at z=0, \(1.03\cdot 10^{11}h^{-1} \) M\( _{\odot } \), corresponds to a
radius, \( R_{vir} = 94.7h^{-1}\) Kpc, which entails 84720 halo particles.
From the accretion history of our halo, we know that it undergoes no
significant merger during the time it hosts our disk, nor immediately before. 
The prolateness of our halo at z=2, where R$_{vir}=$30\,Kpc, is 0.9, 
its spin parameter $\lambda$ has a value 0.04.

\subsection{The baryonic disk}

The spatial distribution of  particles  follows the exponential surface density
law: $\rho =\rho_0\exp -(r/r_0)$ where r$_0$ is the disk scale length,
$r_0=4h^{-1}$\, Kpc, and $\rho_{0}$ is the surface central density. 
The disk is truncated at five scale lengths  with a radius:
R$_{disk}=$20$h^{-1}$\,Kpc. To obtain each disk particle's position according
to the assumed density distribution, we used the rejection method.
We examine two values for the mass of the disk: a more massive case, where
the disk mass is $ 1.9 \times 10^{10}$ solar masses (0.33 in our mass units) 
and a lighter case where the disk mass is   $ 5.9 \times 10^9$ (0.1 in the same units)
The (Plummer equivalent) softening length,  the same for DM, gas, and star particles,
is $ 0.5 h^{-1}$\,Kpc in comoving coordinates.

\section{Simulations}

We performed several cosmological simulations of our disk+halo systems.
The first set (\cite[Curir et al.(2006)]{Cu06})  describes purely stellar
disks, the second set (\cite[Curir et al.(2007)]{Cu07}) 
stellar-gaseous disks with cooling, and the third set simulates the same
stellar-gaseous disks with star formation (\cite[Curir et al.(2008)]{Cu08}).
We exploited the  parallel Tree+SPH N-body code GADGET-2 
(courtesy of V. Springel).
The simulations run on the  CLX computers located at
the CINECA computing centre (BO, Italy ) and on OATo Beowulf-class cluster
of 32 Linux-based PC  at the Osservatorio Astronomico di Torino.

\begin{table}
\begin{center}
\caption{ Simulations: final values.  o.s.=old stars , n.s.= new stars}
{\scriptsize
\begin{tabular}{c c c c c c c c c c }
\hline
N  & M$_{disk}$ & gas fraction & $\epsilon$ (o.s.)&  $\epsilon$ (n.s.)  & a$_{max}$ (o.s.)
& a$_{max}$ (n.s.)&  bulge(o.s.)& bulge(n.s.)  &
bars in bars \\  
 c1 &  0.33 & 0.1 & 0.65 & 0.72 & 8 & 4 & y & n & n \\
 c2 &  0.33 &  0.2 & 0.55 & 0.55 & 11 & 5.7& y& y & n \\
 c3 &  0.33 & 0.4  & 0.6 & 0.55&  8.4 &6& y & y & n \\
 c4 &  0.1 & 0.1 & 0.39 &0.01  & 3 & - & n &y & y \\
 c5 &  0.1 & 0.2 & 0.45 &0.03 & 3  & - & n & y & y\\
 c6 &  0.1 & 0.6 &  0.5 &0.02 &3& - & n & y& y \\ \hline
\end{tabular}
}
\end{center}
\end{table}

The main parameters and the final properties of a set of our simulations, the one where the
star formation is triggered, are listed in Table 1.
The units are $ 5.9 \times 10^{10} M_{\odot }$  for the mass, and
 20 Kpc  for the length.

In the purely stellar disk cases a long living bar (lasting 10 Gyr) appears in
all the simulations.

In our massive stellar-gaseous disk when only the cooling is present, we find  that a
gaseous mass concentration equal to 9\% of the total mass of the disk
inside a radius of 2 Kpc is a lower limit  for the bar dissolution. 
In these  massive disks, where the baryons gravitational field is comparable to that 
of the DM halo, 
we find a threshold value for the gas fraction, 0.2, able to destroy the bar,
whereas in the DM-dominated disks a stellar bar is still leaving at $z=0$
even if the gaseous fraction exceeds  half of the disk mass. 
\begin{figure}
\begin{center}
\includegraphics[width=3.4in]{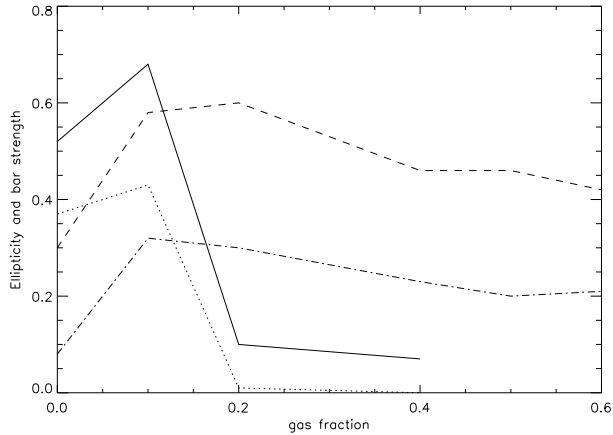}
\caption{ Behaviour of the bar strength and ellipticity at z$=0$ for our set of
cosmological simulations  with increasing gas fraction. 
 $Q_b$ (dotted line) and ellipticity (full line) of our
more massive disks (i.e. disk--to--halo mass ratio
0.33); $Q_b$ (dot--dashed line) and ellipticity (dashed line) of our
less massive, DM-dominated, disks (i.e. disk--to--halo mass ratio
0.1).
}
\label{strengthgas}
\end{center}
\end{figure}
In Fig. 2
we show the behaviour of the ellipticity and of the
 dynamical parameter $Q_b$ (defined by \cite[Combes\& Sanders (1981)]{CoSa} to measure the bar strength) as a function of the increasing mass fraction in disks of 
different masses, where only the cooling is activated. The agreement we recover
between the trends is remarkable: the dynamical evaluation of the bar strength and the geometrical
evaluation of it through ellipticities.

In all our cosmological simulations  with star formation, 
a stellar bar is still living at z=0 in the old star component, even in the
disks having the gas fraction higher than the threshold value for the bar
dissolution discovered in the simulations with gas and cooling.
The new stars component at z=0 is arranged in a bulge component which can
present a barred shape depending on the initial mass of the disk and on
the gaseous fraction (see Fig. 3 and 4).
Fig.4 shows that in the DM-dominated disks the new stars structures appear to have the
characteristics of {\it pseudobulges} as defined by Kormendy, namely of
spheroids formed by secular processes and hiding small  bars inside. 
\begin{figure}
\begin{center}
\includegraphics[width=3.4in]{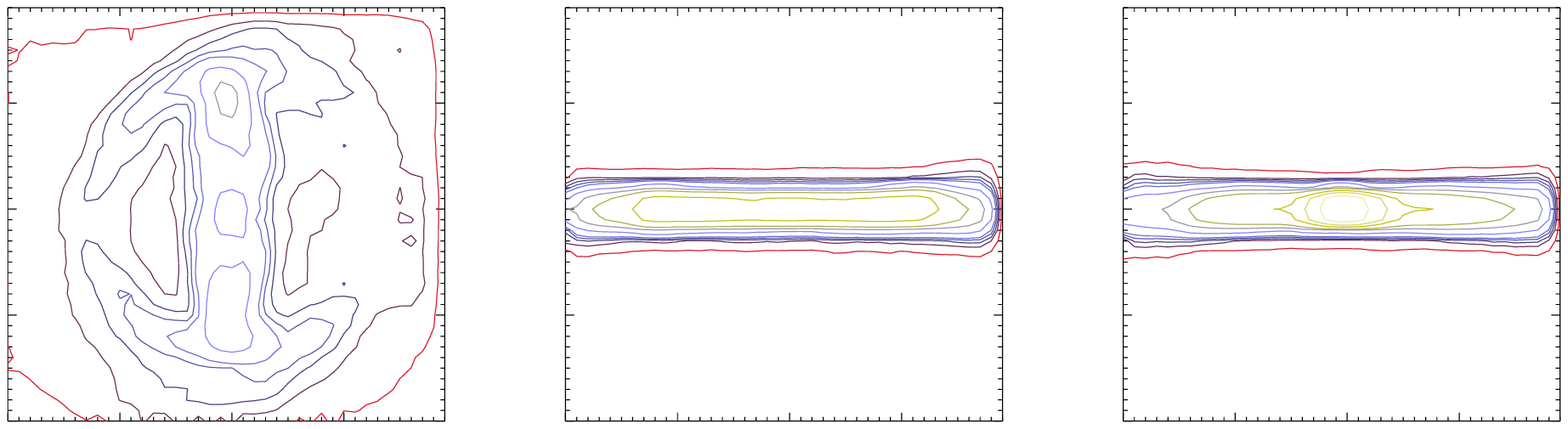}
\includegraphics[width=3.4in]{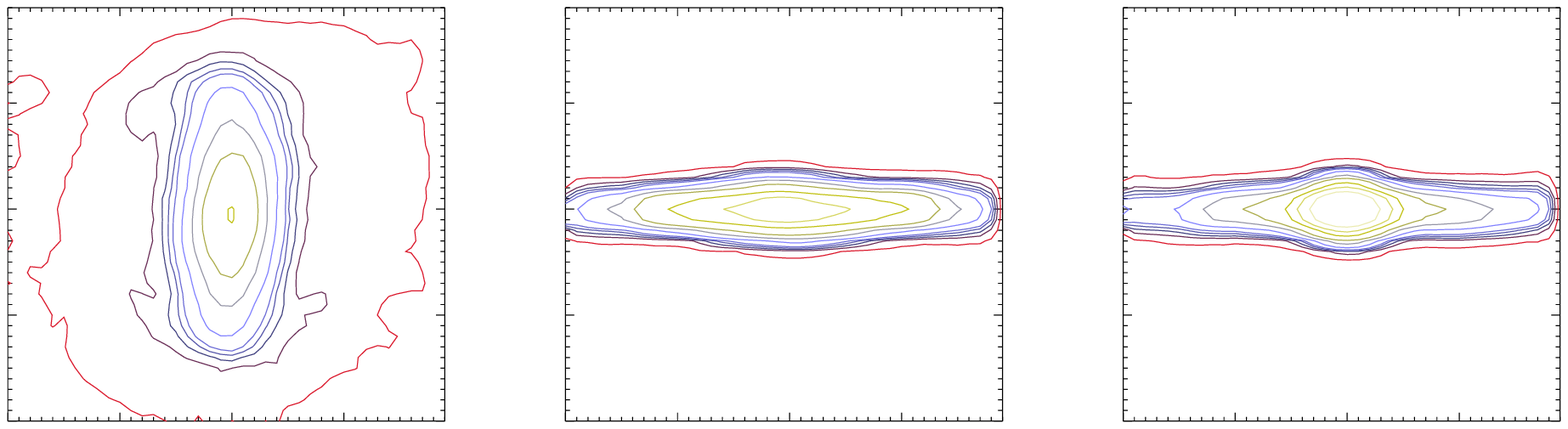}
\includegraphics[width=3.4in]{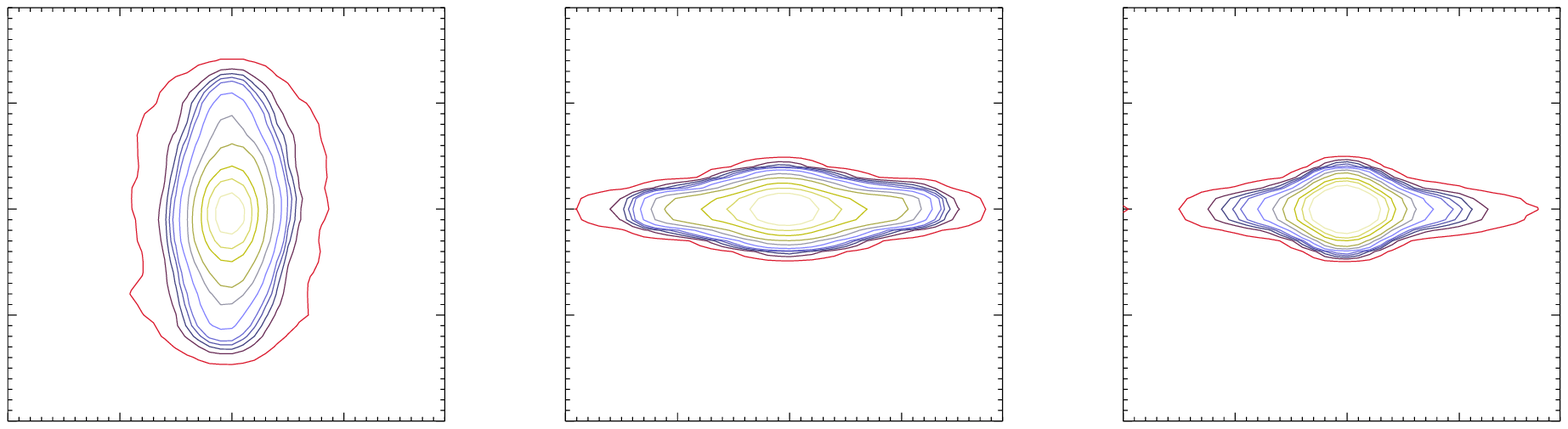}
\caption{Face-on, edge-on, and side-on iso-density contours (from left to right)
of simulation c3 at z=0; top panel shows the gaseous component,
 middle panel the
old star component, bottom panel the newly formed stars.}
\label{bulge}
\end{center}
\end{figure}

\begin{figure}
\begin{center}
\includegraphics[width=3.4in]{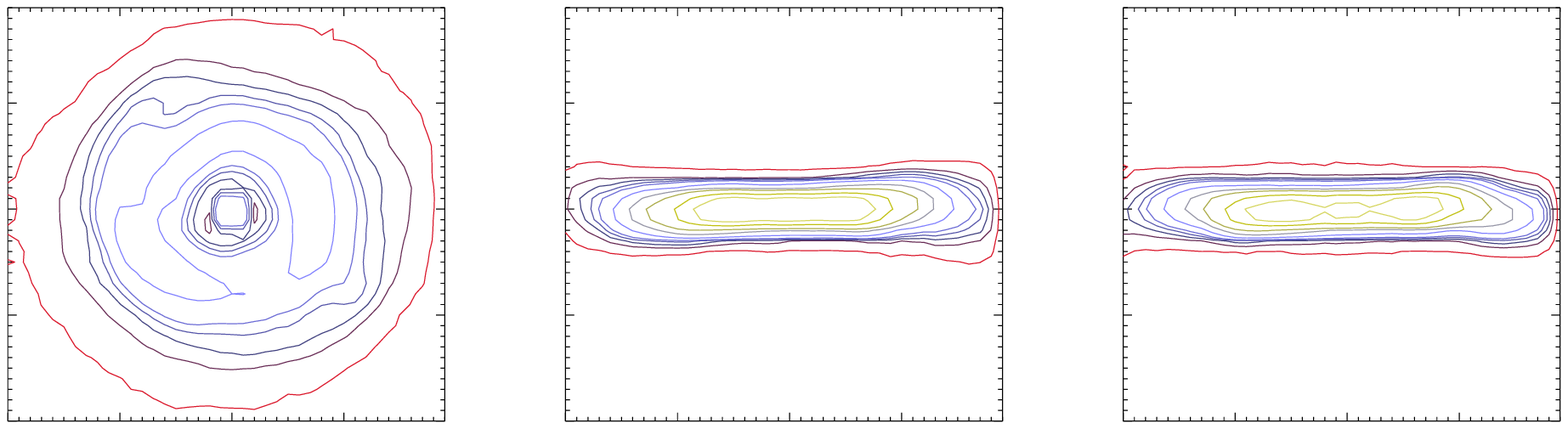}
\includegraphics[width=3.4in]{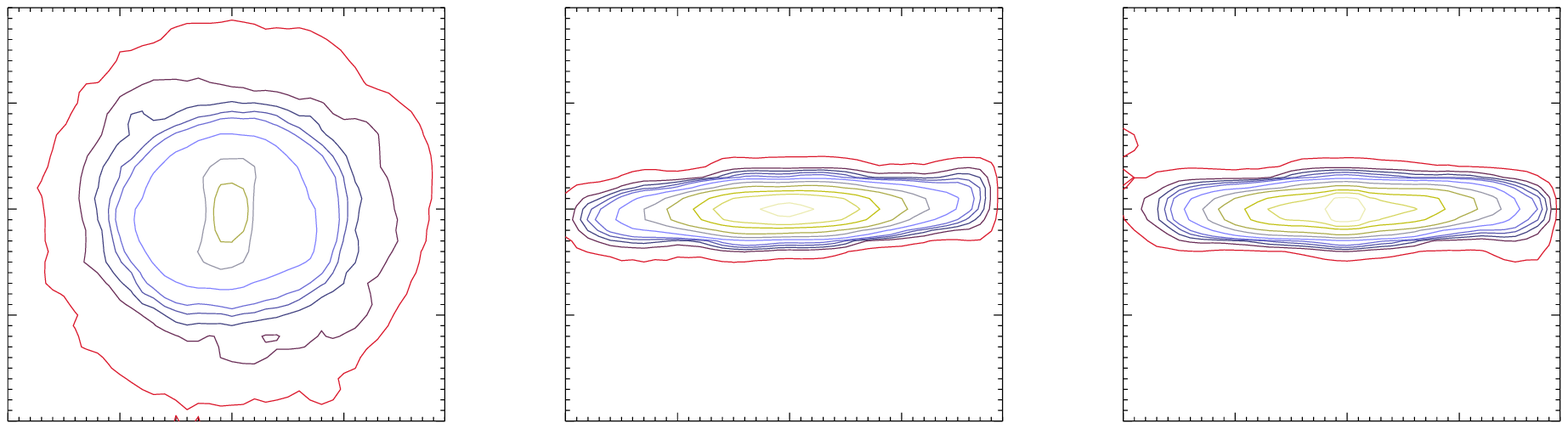}
\includegraphics[width=3.4in]{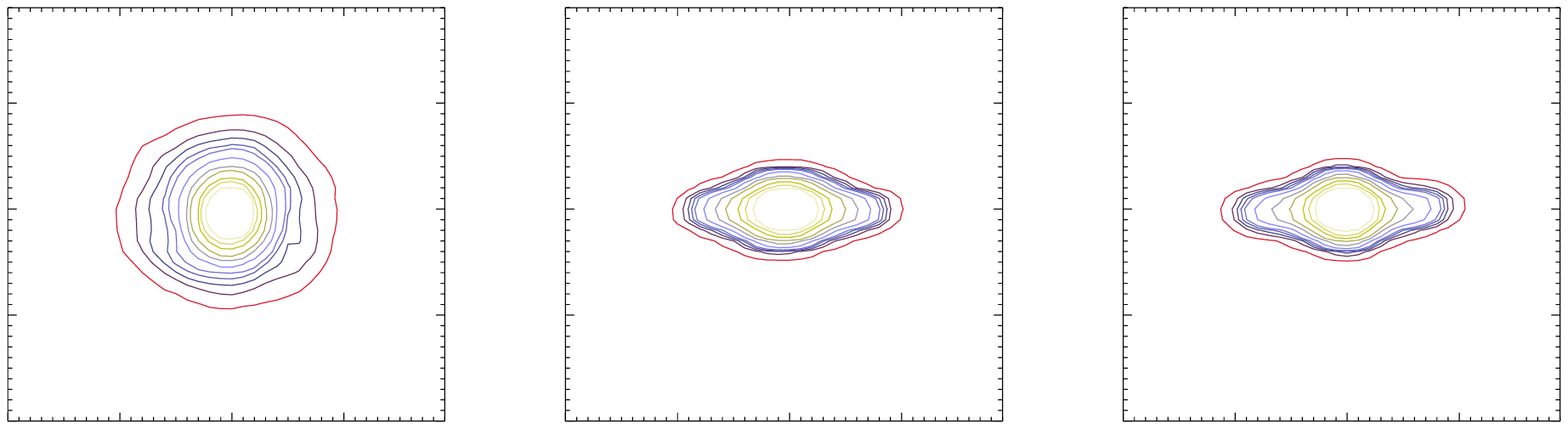}
\caption{Same as in Fig. \ref{bulge}, but for simulation c6}
\label{bulge1}
\end{center}
\end{figure}

In this set  of
  simulations, the  final bar is increasing its
 strength with the increasing gas fraction (Fig. 5). 
\begin{figure}
\begin{center}
\includegraphics[width=3.4in]{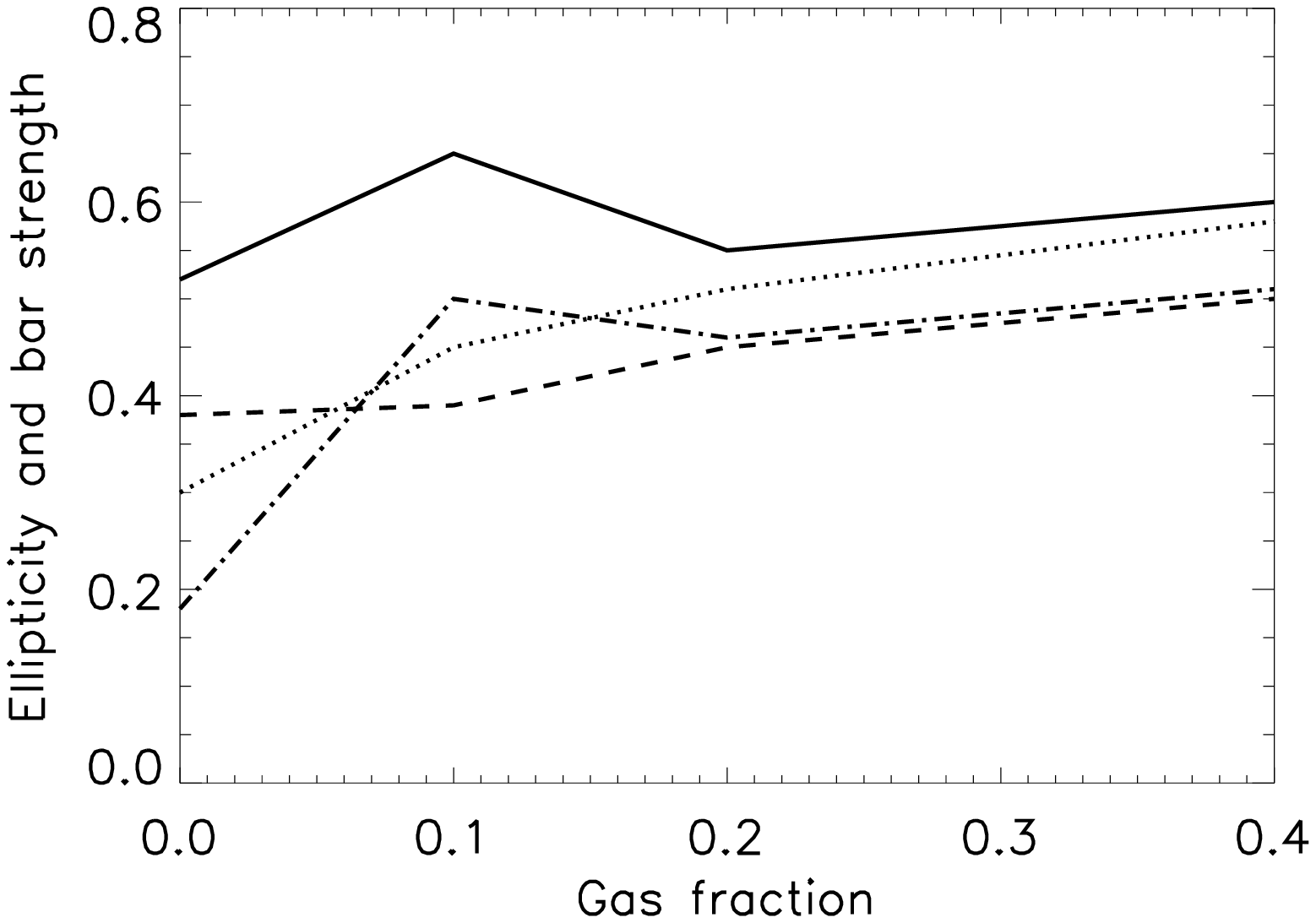}
\end{center}
\caption{ Behaviour of the bar strength and ellipticity at z$=0$ for our set of
cosmological simulations with star formation,  with increasing gas fraction. 
 $Q_b$ (dotted line) and ellipticity (full line) of our
more massive disks (i.e. disk--to--halo mass ratio
0.33); $Q_b$ (dot--dashed line) and ellipticity (dashed line) of our
less massive, DM-dominated, disks (i.e. disk--to--halo mass ratio
0.1).
}
\label{strength}
\end{figure}

\section{Implications}

In \cite[Curir et al.(2006)]{Cu06} we stressed that the  purely stellar DM
dominated disks  show  a behaviour that is strongly driven by the cosmology.
For these disks, when gas and cooling is present 
(\cite[Curir et al. (2007)]{Cu07}) we did not find any  value of
the gas fraction, in the range 0.1--0.6, able to destroy the bar;
moreover even a high value of central gas concentration
does not succeed in dissolving the  bar.  
The bars in
these disks are not a classical product of the self--gravity, which is very
weak, or of  angular momentum exchanges,
since the disk rotates very slowly.  They are
features that strongly  depend on  the dynamical state and  on the
evolution of the cosmological halo, 
therefore the classical
results emphasising the gas impact obtained outside the cosmological scenario are
no longer applicable. 

When the star formation is taken into account (\cite[Curir et al. (2008)]{Cu08}),
the new stars  in DM-dominated disks arrange in pseudobulges
 at z=0,
but a bar  always is maintained
in the old stars population. Also in these  DM-dominated 
disks the bars are not due to classical resonances.  

\section{Conclusions}

In self gravitating disks, cooling gas and star formation produce competitive
effects in the central regions to maintain and enhance (star formation) or to
destroy (gas concentration) the bar feature. In cosmological framework, the
 results for these disks are qualitatively in agreement with the classical ones.
On the other hand for DM-dominated disks the  classical criteria to account for bar instability
cannot be validated in a cosmological framework.
The  mass 
anisotropy, and the dynamical evolution of the DM  cosmological halo have indeed
a crucial effect in enhancing and fuelling the bar instability also in the 
cases where isolated
halo-disk {\it ad hoc} models provided stability predictions.

\end{document}